\documentclass[universe,review,accept,moreauthors,pdftex]{Definitions/mdpi}

\firstpage{1}
\pubvolume{7}
\issuenum{1}
\articlenumber{459}
\pubyear{2021}\copyrightyear{2021}
\externaleditor{{Academic Editor: Susana Cebrian Guajardo, María Martínez Pérez and
Carlos Peña Garay}} 
\datereceived{25 October 2021}
\dateaccepted{20 November 2021}
\datepublished{24 November 2021}
\hreflink{https://\\doi.org/10.3390/universe7120459} 

\usepackage{rotating}

\newcommand{\dCP}{\delta_\text{CP}}
\newcommand{\Dmq}{\Delta m^2}
\newcommand{\eVq}{\ensuremath{\text{eV}^2}}

\makeatletter
\DeclareRobustCommand\recite[1]{\begingroup\@fileswfalse\cite{#1}\endgroup}
\makeatother

\graphicspath{{Figures/}}

\allowdisplaybreaks

\Title{NuFIT: Three-Flavour Global Analyses of Neutrino Oscillation Experiments}

\TitleCitation{NuFIT: Three-Flavour Global Analyses of Neutrino Oscillation Experiments}

\Author{M.C. Gonzalez-Garcia$^{1,2,3,*}$\orcidA{}, Michele Maltoni$^{4,*}$\orcidB{}, Thomas Schwetz$^{5,*}$\orcidC{}}

\AuthorNames{M.C. Gonzalez-Garcia, Michele Maltoni, Thomas Schwetz}

\AuthorCitation{Gonzalez-Garcia, M.C.; Maltoni, M.; Schwetz, T.}

\address{%
  $^1$\quad Instituci\'o Catalana de Recerca i Estudis Avan\c{c}ats (ICREA),
  Pg.\ Lluis Companys 23, 08010 Barcelona, Spain
  \\
  $^2$\quad Departament d'Estructura i Constituents de la Mat\`eria,
  Universitat de Barcelona, 647 Diagonal, E-08028 Barcelona, Spain
  \\
  $^3$\quad C.N.~Yang Institute for Theoretical Physics, SUNY at Stony
  Brook, Stony Brook, NY 11794-3840, USA
  \\
  $^4$\quad Instituto de F\'{\i}sica Te\'orica UAM/CSIC, Calle de
  Nicol\'as Cabrera 13--15, Universidad Aut\'onoma de Madrid,
  Cantoblanco, E-28049 Madrid, Spain
  \\
  $^5$\quad Institut f\"ur Astroteilchenphysik, Karlsruher Institut f\"ur
  Technologie (KIT), D-76021 Karlsruhe, Germany}

\corres{Correspondence: concha@insti.physics.sunysb.edu (M.C.G.-G.); 
  michele.maltoni@csic.es (M.M.) schwetz@kit.edu (T.S.)}

\abstract{In this contribution, we summarise the determination of
  neutrino masses and mixing arising from  global analysis of data
  from atmospheric, solar, reactor, and accelerator neutrino
  experiments performed in the framework of three-neutrino mixing and
  obtained in the context of the NuFIT collaboration.  Apart from
  presenting the latest status as of autumn 2021, we discuss the
  evolution of  global-fit results over the last 10 years, and
  mention various pending issues (and their resolution) that occurred
  during that period in the global analyses.}


\begin{document}

\section{Introduction}
\label{sec:intro}

The observation of flavour transitions in neutrino propagation in a
variety of experiments has established beyond doubt that lepton
flavours are not symmetries of nature. The dependence of the
probability of  observed flavour transitions with the distance
travelled by the neutrinos and their energy has allowed for singling out
neutrino masses and the mixing in  weak charged current
interactions of the massive neutrino states as the responsible
mechanism for the observed flavour
oscillations~\cite{Pontecorvo:1967fh, Gribov:1968kq} (see~\cite{GonzalezGarcia:2007ib} for an~overview).

At the time of  writing  this minireview, neutrino oscillation
effects have been observed in:
\begin{itemize}
\item $\nu_e$, $\nu_\mu$, $\bar\nu_e$, and $\bar \nu_\mu$ atmospheric
  neutrinos, produced by the interaction of the cosmic rays on the top
  of the atmosphere. Results with the highest statistics
  correspond to  Super-Kamiokande~\cite{Abe:2017aap} and
  IceCube/DeepCore~\cite{Aartsen:2014yll, deepcore:2016} experiments.

\item $\nu_e$ solar neutrinos produced in  nuclear reactions that
  make the Sun shine. Results included in the present determination of
  the flavour evolution of solar neutrinos comprise the total event
  rates in radiochemical experiments Chlorine~\cite{Cleveland:1998nv},
  Gallex/GNO~\cite{Kaether:2010ag}, and
  SAGE~\cite{Abdurashitov:2009tn}, and the time- and energy-dependent
  rates in the four phases of Super-Kamiokande~\cite{Hosaka:2005um,
    Cravens:2008aa, Abe:2010hy, SK:nu2020}, the three phases of
  SNO~\cite{Aharmim:2011vm}, and Borexino~\cite{Bellini:2011rx,
    Bellini:2008mr, Bellini:2014uqa}.

\item neutrinos produced in accelerators and detected at distance
  $\mathcal{O}(\text{100~km})$, in the so-called long baseline (LBL)
  experiments, and in which neutrino oscillations have been observed
  in two channels:
  \begin{itemize}
  \item disappearance results in the energy distribution of $\nu_\mu$
    and $\bar\nu_\mu$ events that were  precisely measured in
    MINOS~\cite{Adamson:2013whj}, T2K~\cite{T2K:nu2020}, and
    NOvA~\cite{NOvA:nu2020}.
    
  \item appearance results of both $\nu_e$ and $\bar\nu_e$ events 
    in their energy distribution detected in
    MINOS~\cite{Adamson:2013ue}, T2K~\cite{T2K:nu2020}, and
    NOvA~\cite{NOvA:nu2020}.
  \end{itemize}
  
\item $\bar\nu_e$ produced in nuclear reactors. Their disappearance
was  observed in their measured energy spectrum at two
  distinctive baselines.
  \begin{itemize}
  \item at $\mathcal{O}(\text{1~km})$,  denoted medium baselines
    (MBL), in Double Chooz~\cite{DoubleC:nu2020}, Daya
    Bay~\cite{Adey:2018zwh}, and RENO~\cite{RENO:nu2020}.

  \item at LBL in KamLAND~\cite{Gando:2013nba}.
  \end{itemize}
\end{itemize}

These results imply that neutrinos are
massive and there is physics beyond the Standard Model (BSM). 

The first step towards the discovery of the underlying BSM dynamics
for neutrino masses is the detailed characterisation of the minimal
low-energy parametrisation  that can describe the bulk of results.
This requires global analysis of  oscillation data as they become
available. At present, such combined analyses are in the hands of a
few phenomenological groups (see, for example,~\cite{deSalas:2020pgw,deSalas:2018bym, Capozzi:2021fjo,Capozzi:2020qhw}). Results obtained
by the different groups are generically in good agreement, which provides
a test of the robustness of the present determination of the oscillation parameters. 
The NuFIT Collaboration~\cite{nufit} was formed in this context about one decade
ago by the three authors of this article with the goal of providing
timely updated global analysis of neutrino oscillation measurements
determining the leptonic mixing matrix and the neutrino masses in the
framework of the Standard Model extended with three massive neutrinos.
We  published five major updates of the
analysis~\cite{GonzalezGarcia:2012sz, Gonzalez-Garcia:2014bfa,
  Bergstrom:2015rba, Esteban:2016qun, Esteban:2018azc,
  Esteban:2020cvm}, while intermediate updates are  regularly
posted in the NuFIT website~\cite{nufit}. Over the years, the work of a
number of graduate students and postdocs has been paramount to the
success of the project: Johannes Bergström~\cite{Bergstrom:2015rba},
Ivan Esteban~\cite{Esteban:2016qun, Esteban:2018azc, Esteban:2020cvm},
Alvaro Hernandez-Cabezudo~\cite{Esteban:2018azc}, Ivan
Martinez-Soler~\cite{Esteban:2016qun}, Jordi
Salvado~\cite{GonzalezGarcia:2012sz}, and Albert
Zhou~\cite{Esteban:2020cvm}.

\section{ New Minimal Standard Model with Three Massive Neutrinos}
\label{sec:frame}

The Standard Model (SM) is a gauge theory
built to explain the strong, weak, and electromagnetic interactions of
all known elementary particles. It is based on  gauge symmetry
$SU(3)_\text{Color} \times SU(2)_\text{Left}\times U(1)_\text{Y}$ and
 is spontaneously broken to $SU(3)_\text{Color}\times
U(1)_\text{EM}$ by the Higgs mechanism, which provides a vacuum expectation
value for a Higgs  $SU(2)_\text{Left}$ doublet field $\phi$.  The SM contains three fermion
generations and the chiral nature of the $SU(2)_\text{Left}$ part of
the gauge group that is partly responsible for the weak interactions implies that right- and left-handed fermions experience
different weak interactions.  Left-handed fermions are assigned to
the $SU(2)_\text{Left}$ doublet representation, while right-handed fermions
are $SU(2)_\text{Left}$ singlets.  Because of the vector
nature of $SU(3)_\text{Color}$ and $U(1)_\text{EM}$ interactions, both
left- and right-handed fermion fields are required to build 
electromagnetic and strong currents. Neutrinos are the only
fermions that   have neither color nor electric charge. They
only feel weak interactions. Consequently,  right-handed neutrinos
are singlets of the full SM group and thereby have no place in the
SM of particle interactions.

As a consequence of the gauge symmetry and  group representations
in which  fermions are assigned, the SM possesses  accidental
global symmetry $U(1)_B \times U(1)_e \times U(1)_\mu \times
U(1)_\tau$, where $U(1)_B$ is the baryon number symmetry, and
$U(1)_{e,\mu,\tau}$ are the three lepton flavour symmetries.

In the SM, fermion masses are generated by  Yukawa interactions
that couple the right-handed fermion ($SU(2)_\text{Left}$-singlet) to
the left-handed fermion ($SU(2)_\text{Left}$-doublet) and the Higgs
doublet.  After electroweak spontaneous symmetry breaking, these
interactions provide charged fermion masses. No Yukawa
interaction can be written that would give mass to the neutrino
because no right-handed neutrino exists in the model.  Furthermore,
any neutrino mass term built with the left-handed neutrino fields
would violate $U(1)_{L=L_e+L_\mu+L_\tau}$, which is a subgroup of the
accidental symmetry group. As such, it cannot be generated by loop
corrections within the model.  It can also not be generated by
nonperturbative corrections because the $U(1)_{B-L}$ subgroup of the
global symmetry is nonanomalous.

From these arguments, it follows that the SM predicts that neutrinos
are strictly massless. It also implies that there is no
leptonic flavour mixing, and that there is no possibility of CP
violation of the leptons. However, as described in the introduction, we
have now undoubted experimental evidence that leptonic flavours are not
conserved in neutrino propagation. The Standard Model must thus
 be extended.

The simplest extension capable of describing the experimental
observations must include neutrino masses.  Let us call this minimal
extension the Minimally Extended Standard Model (NMSM). In fact, this simplest
extension is not unique because, unlike for charged fermions, one can
construct a neutrino mass term in two different forms:
\begin{itemize}
\item in one minimal extension,    right-handed
  neutrinos $\nu_R$ are introduced, and  that the total lepton number is
  still conserved is imposed. In this form, gauge invariance allows for a Yukawa
  interaction involving $\nu_R$. and the lepton doublet, in analogy to the charged
  fermions, after electroweak spontaneous symmetry breaking the
  NMSM Lagrangian, reads:
  \begin{equation}
    \label{eq:dirac}
    \mathcal{L}_D = \mathcal{L}_\text{SM}
    - M_\nu \bar \nu_L \nu_R + \text{h.c.}
  \end{equation}
  In this case, the neutrino mass eigenstates are Dirac fermions, and
  neutrino and antineutrinos are distinct fields, {i.e.},
  $\nu^c \neq \nu$ ($\nu^C$ here represents the charge conjugate
  neutrino field). This NMSM is gauge-invariant under the SM gauge group.

\item In another minimal extension,   a mass term is constructed
  employing only the SM left-handed neutrinos by allowing for the
  violation of the total lepton number. In this case, the NMSM Lagrangian
  is
  \begin{equation}
    \mathcal{L}_M = \mathcal{L}_\text{SM}
    -\frac{1}{2} M_\nu \bar \nu_L \nu_L^c + \text{h.c.}
  \end{equation}
  In this NMSM, mass eigenstates are Majorana fermions, $\nu^c =
  \nu$. The Majorana mass term above breaks  electroweak gauge
  invariance.
  \end{itemize}

Consequently, $\mathcal{L}_M$ can only be understood as a low-energy
limit of a complete theory, while $\mathcal{L}_D$ is formally
self-consistent.  However, in either NMSM, lepton flavours are mixed into the
charged-current interactions of the leptons.  We denote
the neutrino mass eigenstates
by $\nu_i$ with $i=1,2,\ldots$, and the charged lepton mass
eigenstates by $l_i=(e,\mu,\tau)$; then, the leptonic charged-current
interactions of those massive states are given by
\begin{equation}
  \label{CClepmas}
  -\mathcal{L}_\text{CC} = \frac{g}{\sqrt{2}}\, \overline{l_{iL}}\,
  \gamma^\mu \, U^{ij}\, \nu_j\, W_\mu^- + \text{h.c.}
\end{equation}
where $U$ is the leptonic mixing matrix analogous to the CKM matrix
for the quarks. Leptonic mixing generated this way is, however,
slightly more general than the CKM flavour mixing of quarks because the
number of massive neutrinos ($n$) is unknown. This is so
because right-handed neutrinos are SM singlets; therefore, there are
no constraints on their number. As mentioned above, unlike
charged fermions, neutrinos can be Majorana particles.  As a
consequence, the number of new parameters in the model NMSM depends on
the number of massive neutrino states and on whether they are Dirac or
Majorana particles.

In most generality $U$ in Equation~\eqref{CClepmas} is a $3\times n$ matrix, and
verifies $U U^\dagger =I_{3\times 3}$ but in general $U^\dagger U \neq
I_{n\times n}$. In this review, however, we   focus
on analyses made in the context of only three neutrino massive states,
which is the simplest scheme to consistently describe  data listed
in the introduction. In this case, the three known neutrinos ($\nu_e$,
$\nu_\mu$, $\nu_\tau$) can be expressed as quantum superpositions of
three massive states $\nu_i$ ($i=1,2,3$) with masses $m_i$, and the
leptonic mixing matrix can be parametrized as~\cite{PDG}:
\begin{equation}
  \label{eq:matrix}
  U =\begin{pmatrix}
  1 & 0 & 0 \\
  0 & c_{23}  & {s_{23}} \\
  0 & -s_{23} & {c_{23}}
  \end{pmatrix}
  \begin{pmatrix}
    c_{13} & 0 & s_{13} e^{-i\dCP} \\
    0 & 1 & 0 \\
    -s_{13} e^{i\dCP} & 0 & c_{13}
  \end{pmatrix}
  \begin{pmatrix}
    c_{12} & s_{12} & 0 \\
    -s_{12} & c_{12} & 0 \\
    0 & 0 & 1
  \end{pmatrix}
  \begin{pmatrix}
    e^{i \alpha_1} & 0 & 0 \\
    0 & e^{i \alpha_2} & 0 \\
    0 & 0 & 1
  \end{pmatrix},
\end{equation}
where $c_{ij} \equiv \cos\theta_{ij}$ and $s_{ij} \equiv
\sin\theta_{ij}$.  In addition to  Dirac-type phase $\dCP$,
analogous to that of the quark sector, there are two physical phases,
$\alpha_1$ and $\alpha_2$, associated to a possible Majorana character of
neutrinos that, however, are not relevant for neutrino oscillations.

A consequence of the presence of neutrino masses and the leptonic
mixing is the possibility of mass-induced flavour oscillations in
vacuum~\cite{Pontecorvo:1967fh, Gribov:1968kq}, and of flavour transitions when a neutrino traverses regions of
dense matter~\cite{Wolfenstein:1977ue, Mikheev:1986gs}. Generically, the
flavour transition probability in vacuum presents an oscillatory $L$
dependence with phases proportional to $\sim \Delta m^2 L/E$ and
amplitudes proportional to different elements of mixing matrix.  The
presence of matter in the neutrino propagation alters both the
oscillation frequencies and the amplitudes (see~\cite{GonzalezGarcia:2007ib} for an overview).

In the convention of Equation~\eqref{eq:matrix}, the disappearance of solar
$\nu_e$ and long baseline reactor $\bar\nu_e$ dominantly proceeds
via oscillations with wavelength $\propto E / \Dmq_{21}$ ($\Dmq_{ij}
\equiv m_i^2 - m_j^2$ and $\Dmq_{21} \geq 0$ by convention) and
amplitudes controlled by $\theta_{12}$, while the disappearance of
atmospheric and LBL accelerator $\nu_\mu$ dominantly proceeds via
oscillations with wavelength $\propto E / |\Dmq_{31}| \ll E/\Dmq_{21}$
and amplitudes controlled by $\theta_{23}$. Generically, $\theta_{13}$
controls the amplitude of oscillations involving $\nu_e$ flavour with
$E/|\Dmq_{31}|$ wavelengths.  Angles $\theta_{ij}$ can be taken to
lie in the first quadrant, $\theta_{ij} \in [0,\pi/2]$, and  phase
$\dCP \in [0, 2\pi]$.  Values of $\dCP$ different from 0 and $\pi$
imply CP violation in neutrino oscillations in vacuum. In this
convention, $\Delta m^2_{21}$ is positive by construction.  Moreover,
given the observed hierarchy between the solar and atmospheric
wavelengths, there are two possible nonequivalent orderings for the
mass eigenvalues:
\begin{itemize}
\item $m_1\ll m_2< m_3$ so $\Dmq_{21} \ll \Dmq_{32} (\simeq \Dmq_{31}
  > 0) $, referred to as Normal Ordering (NO);

\item $m_3\ll m_1< m_2$ so $\Dmq_{21} \ll -(\Dmq_{31} \simeq \Dmq_{32}
  < 0)$ referred to as Inverted Ordering (IO).
\end{itemize}

The two orderings, therefore, correspond to the two possible choices of
the sign of $\Dmq_{31}$. In NuFIT, we adopted the convention of
reporting results for $\Dmq_{31}$ for NO and $\Dmq_{32}$ for IO,
{i.e.}, we always use the one that has the larger absolute
value. We sometimes generically denote such quantity as
$\Dmq_{3\ell}$, with $\ell=1$ for NO and $\ell=2$ for IO.

In summary,  $3\nu$ oscillation analysis of the existing data
involves a total of six parameters: two mass-squared differences (one
of which can be positive or negative), three mixing angles, and  CP
phase $\dCP$.  For the sake of clarity, we summarise which experiment contributes dominantly to the
present determination of the different~parameters in
Table~\ref{tab:expe}.

\begin{specialtable}[H]
  \caption{Experiments contributing to the present determination of
     oscillation parameters.}
  \begin{small}
  \setlength{\tabcolsep}{3.15mm}
    \begin{tabular}{lll}
 \toprule    
     \textbf{ Experiment} & \textbf{Dominant }& \textbf{Important}
      \\
      \midrule
      Solar Experiments
      & $\theta_{12}$
      & $\Dmq_{21}$, $\theta_{13}$
      \\
      Reactor LBL (KamLAND)
      & $\Dmq_{21}$
      & $\theta_{12}$, $\theta_{13}$
      \\
      Reactor MBL (Daya Bay, RENO, Double Chooz)
      & $\theta_{13}$, $|\Dmq_{3\ell}|$
      & ---
      \\
      Atmospheric Experiments (SK, IC-DC)
      & ---
      & $\theta_{23}$, $|\Dmq_{3\ell}|$, $\theta_{13}$, $\dCP$
      \\
      Accel. LBL $(\nu_\mu, \bar\nu_\mu)$ disapp. (K2K, MINOS, T2K, NOvA)
      & $|\Dmq_{3\ell}|$, $\theta_{23}$
      & ---
      \\
      Accel. LBL $(\nu_e, \bar\nu_e)$ appearance (MINOS, T2K, NOvA)
      & $\dCP$
      & $\theta_{13}$, $\theta_{23}$
      \\
      \bottomrule
    \end{tabular}
  \end{small}
  \label{tab:expe}
\end{specialtable}

\section{NuFIT Results: The Three-Neutrino Paradigm}
\label{sec:nufit}

\begin{figure}[H]
\widefigure
  \includegraphics[width=\linewidth]{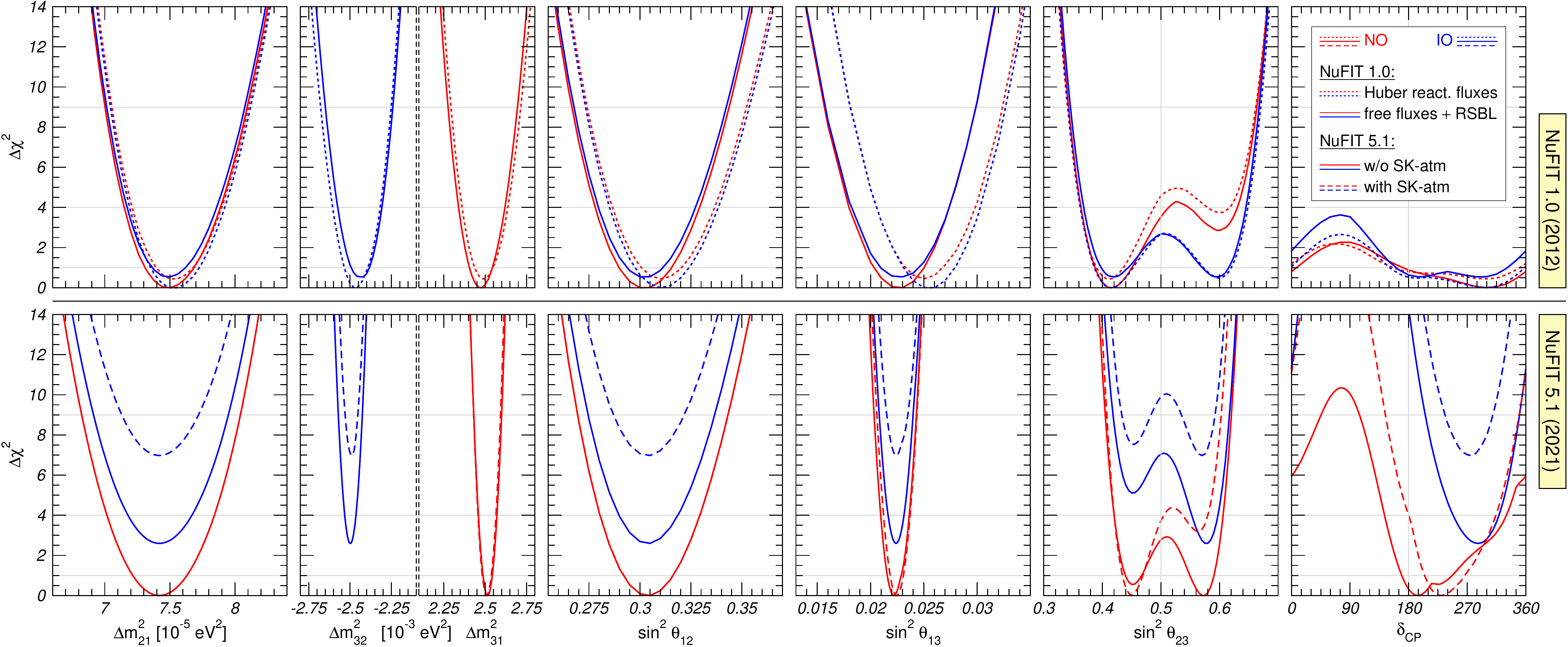}\hfill
  \caption{{Comparison} of  global $3\nu$ oscillation analysis
    results.  All panels  show $\Delta\chi^2$ profiles minimised
    with respect to all undisplayed parameters.  Red (blue) curves
    are for normal (inverted) ordering.  As atmospheric
    mass-squared splitting, we used $\Dmq_{31}$ for NO and $\Dmq_{32}$
    for IO. \textbf{(top)} NuFIT~1.0 results. Solid curves  obtained
    with free normalisation of reactor fluxes and the inclusion of
    data from short-baseline (less than 100~m) reactor experiments;
     for  dotted curves, short-baseline data were not
    included, but reactor fluxes were fixed to the predictions of~\cite{Huber:2011wv}. \textbf{(bottom)}~NuFIT~5.1 results. In
    all curves, theneutrino fluxes for each reactor experiment were
    constrained by the corresponding near detector.  Solid (dashed)
    curves  obtained without (with)  including  tabulated
    SK-atm $\Delta\chi^2$ data. Figure adapted from~\cite{nufit}.}
  \label{fig:compachis}
\end{figure}

The latest determination of the six parameters in the new NMSM is
presented in Table~\ref{tab:bfranges}, corresponding to 
NuFIT~5.1 analysis~\cite{Esteban:2020cvm, nufit}.  Progress in the
determination of these parameters over the last decade is illustrated
in Figure~\ref{fig:compachis}, which shows the one-dimensional
projections of the $\Delta\chi^2$ from  global analysis as a
function of each of the six parameters, obtained in the first NuFIT
1.0 analysis and the last NuFIT~5.1 in the upper and lower rows,
respectively.

\begin{specialtable}[H]
  \caption{Determination of  three-flavour oscillation
    parameters from  fit to global data
    NuFIT~5.1~\recite{Esteban:2020cvm, nufit}.  Results in the
    first and second columns correspond to  analysis performed under
    the assumption of NO and IO, respectively; therefore, they are
    confidence intervals defined relative to the respective local
    minimum.  Results shown in the upper and lower sections
    correspond to analysis performed without and with the addition of
     tabulated SK-atm $\Delta\chi^2$ data respectively.  In quoting
     values for the largest mass splitting, we  defined
    $\Dmq_{3\ell} \equiv \Dmq_{31} > 0$ for NO and $\Dmq_{3\ell}
    \equiv \Dmq_{32} < 0$ for IO.}
  \begin{small}
  \setlength{\tabcolsep}{2.35mm}
    \begin{tabular}{clcccc}
      \toprule
      \multirow{11}{*}{\begin{sideways}\hspace*{-7em}without SK atmospheric data\end{sideways}} &
      & \multicolumn{2}{c}{Normal Ordering (Best Fit)}
      & \multicolumn{2}{c}{Inverted Ordering ($\Delta\chi^2=2.6$)}
      \\
      \cline{3-6}
      && bfp $\pm 1\sigma$ & $3\sigma$ Range
      & bfp $\pm 1\sigma$ & $3\sigma$ Range
      \\
      \cline{2-6}
      \rule{0pt}{4mm}\ignorespaces
      & $\sin^2\theta_{12}$
      & $0.304_{-0.012}^{+0.013}$ & $0.269 \to 0.343$
      & $0.304_{-0.012}^{+0.012}$ & $0.269 \to 0.343$
      \\[1mm]
      & $\theta_{12}/^\circ$
      & $33.44_{-0.74}^{+0.77}$ & $31.27 \to 35.86$
      & $33.45_{-0.74}^{+0.77}$ & $31.27 \to 35.87$
      \\[3mm]
      & $\sin^2\theta_{23}$
      & $0.573_{-0.023}^{+0.018}$ & $0.405 \to 0.620$
      & $0.578_{-0.021}^{+0.017}$ & $0.410 \to 0.623$
      \\[1mm]
      & $\theta_{23}/^\circ$
      & $49.2_{-1.3}^{+1.0}$ & $39.5 \to 52.0$
      & $49.5_{-1.2}^{+1.0}$ & $39.8 \to 52.1$
      \\[3mm]
      & $\sin^2\theta_{13}$
      & $0.02220_{-0.00062}^{+0.00068}$ & $0.02034 \to 0.02430$
      & $0.02238_{-0.00062}^{+0.00064}$ & $0.02053 \to 0.02434$
      \\[1mm]
      & $\theta_{13}/^\circ$
      & $8.57_{-0.12}^{+0.13}$ & $8.20 \to 8.97$
      & $8.60_{-0.12}^{+0.12}$ & $8.24 \to 8.98$
      \\[3mm]
      & $\dCP/^\circ$
      & $194_{-25}^{+52}$ & $105 \to 405$
      & $287_{-32}^{+27}$ & $192 \to 361$
      \\[3mm]
      & $\dfrac{\Dmq_{21}}{10^{-5}~\eVq}$
      & $7.42_{-0.20}^{+0.21}$ & $6.82 \to 8.04$
      & $7.42_{-0.20}^{+0.21}$ & $6.82 \to 8.04$
      \\[3mm]
      & $\dfrac{\Dmq_{3\ell}}{10^{-3}~\eVq}$
      & $+2.515_{-0.028}^{+0.028}$ & $+2.431 \to +2.599$
      & $-2.498_{-0.029}^{+0.028}$ & $-2.584 \to -2.413$
      \\[2mm]
    \midrule 
      \multirow{11}*{\begin{sideways}\hspace*{-7em}with SK atmospheric data\end{sideways}} &
      & \multicolumn{2}{c}{Normal Ordering (Best Fit)}
      & \multicolumn{2}{c}{Inverted Ordering ($\Delta\chi^2=7.0$)}
      \\
      \cline{3-6}
      && bfp $\pm 1\sigma$ & $3\sigma$ range
      & bfp $\pm 1\sigma$ & $3\sigma$ range
      \\
      \cline{2-6}
      \rule{0pt}{4mm}\ignorespaces
      & $\sin^2\theta_{12}$
      & $0.304_{-0.012}^{+0.012}$ & $0.269 \to 0.343$
      & $0.304_{-0.012}^{+0.013}$ & $0.269 \to 0.343$
      \\[1mm]
      & $\theta_{12}/^\circ$
      & $33.45_{-0.75}^{+0.77}$ & $31.27 \to 35.87$
      & $33.45_{-0.75}^{+0.78}$ & $31.27 \to 35.87$
      \\[3mm]
      & $\sin^2\theta_{23}$
      & $0.450_{-0.016}^{+0.019}$ & $0.408 \to 0.603$
      & $0.570_{-0.022}^{+0.016}$ & $0.410 \to 0.613$
      \\[1mm]
      & $\theta_{23}/^\circ$
      & $42.1_{-0.9}^{+1.1}$ & $39.7 \to 50.9$
      & $49.0_{-1.3}^{+0.9}$ & $39.8 \to 51.6$
      \\[3mm]
      & $\sin^2\theta_{13}$
      & $0.02246_{-0.00062}^{+0.00062}$ & $0.02060 \to 0.02435$
      & $0.02241_{-0.00062}^{+0.00074}$ & $0.02055 \to 0.02457$
      \\[1mm]
      & $\theta_{13}/^\circ$
      & $8.62_{-0.12}^{+0.12}$ & $8.25 \to 8.98$
      & $8.61_{-0.12}^{+0.14}$ & $8.24 \to 9.02$
      \\[3mm]
      & $\dCP/^\circ$
      & $230_{-25}^{+36}$ & $144 \to 350$
      & $278_{-30}^{+22}$ & $194 \to 345$
      \\[3mm]
      & $\dfrac{\Dmq_{21}}{10^{-5}~\eVq}$
      & $7.42_{-0.20}^{+0.21}$ & $6.82 \to 8.04$
      & $7.42_{-0.20}^{+0.21}$ & $6.82 \to 8.04$
      \\[3mm]
      & $\dfrac{\Dmq_{3\ell}}{10^{-3}~\eVq}$
      & $+2.510_{-0.027}^{+0.027}$ & $+2.430 \to +2.593$
      & $-2.490_{-0.028}^{+0.026}$ & $-2.574 \to -2.410$
      \\[2mm]
      \bottomrule
    \end{tabular}
  \end{small}
  \label{tab:bfranges}
\end{specialtable}

To further illustrate the improvement on the robust precision on the
determination of these parameters over the last decade, we could compute
the $3\sigma$ relative precision of parameter $x$
$$ \frac{2(x^{+} - x^{-})}{(x^{+} + x^{-})}$$
where $x^\text{+}$ and $x^{-}$ are the upper and lower bounds on
parameter $x$ at the $3\sigma$ level. Doing so, we find the following
change in the $3\sigma$ relative precision (marginalising over
ordering):
\vspace{6pt}
\begin{equation}
  \label{eq:compaprec}
  \catcode`?=\active\def?{\hphantom{0}}
  \catcode`!=\active\def!{\hphantom{.}}
  \begin{array}{rccccc}
    & \text{NuFIT~1.0} & \text{NuFIT~2.0} & \text{NuFIT~3.0}
    & \text{NuFIT~4.0} &\text{NuFIT~5.1}
    \\
    \hline
    \theta_{12}   & ?15\% & ?14\% & ?14\% & !14\% & !14\% \\
    \theta_{13}   & ?30\% & ?15\% & ?11\% & 8.9\% & 9.0\% \\
    \theta_{23}   & ?43\% & ?32\% & ?32\% & !27\% & !27\% \\
    \Dmq_{21}     & ?14\% & ?14\% & ?14\% & !16\% & !16\%  \\
    |\Dmq_{3\ell}| & ?17\% & ?11\% & ??9\% & 7.8\% & 6.7\%~[6.5\%] \\
    \dCP  &  100\% & 100\% & 100\% & 100\%~[92\%] & 100\%~[83\%] \\
    \Delta\chi^2_\text{IO-NO}
    & \pm 0.5 & -0.97 & +0.83 & +4.7~[+9.3] & +2.6~[+7.0]
  \end{array}
\end{equation}
In the last two columns,  numbers between brackets show the
impact of including  tabulated SK-atm data (see
Section~\ref{sec:skatm}) in the precision of the determination of such a
parameter. Since the $\Delta\chi^2$ profile of $\dCP$ is not Gaussian,
the precision estimation above for $\dCP$ is only indicative. In
addition, the last line shows the $\Delta\chi^2$ between orderings
that, for NuFIT~1.0, changed from $+0.5$ to $-0.5$ depending on the
choice of normalisation for the reactor fluxes.

\end{paracol}
\nointerlineskip
\begin{paracol}{2}
\switchcolumn

Besides the expected improvement on the precision associated with the
increased statistics of some of the experiments and the addition of
data from new experiments, there were a number of issues entering 
analysis, which changed over the covered period. Next, we briefly
comment on those.

\subsection{Reactor Neutrino Flux Uncertainties}
\label{sec:reaflux}

The NuFIT~1.0 analysis, which was conducted soon after the first results from the
medium baseline ($\mathcal{O}(\text{1~km})$) reactor experiments Daya
Bay~\cite{An:2012eh}, RENO~\cite{Ahn:2012nd}, and Double
Chooz~\cite{Abe:2011fz}, provided a positive determination of 
mixing angle $\theta_{13}$. Data from those experiments were
analysed  with those from  finalised reactor experiments
Chooz~\cite{Apollonio:1999ae} and Palo Verde~\cite{Piepke:2002ju}.
Analysis of reactor experiments without a near detector, in
particular Chooz, Palo Verde and the early measurements of Double
Chooz, depends on the expected rates as computed with some prediction
for the neutrino fluxes from the reactors.

At about the same time, the so-called \textit{{reactor} anomaly} was 
first pointed out. It amounted to the fact that the most updated reactor
flux calculations in~\cite{Mueller:2011nm, Huber:2011wv,Mention:2011rk} resulted in an increase in the predicted fluxes and
a reduction in  uncertainties. Compared to those fluxes,  results
from finalised reactor experiments at baselines $\lesssim$100~m such as
Bugey4~\cite{Declais:1994ma}, ROVNO4~\cite{Kuvshinnikov:1990ry},
Bugey3~\cite{Declais:1994su}, Krasnoyarsk~\cite{Vidyakin:1987ue,
  Vidyakin:1994ut}, ILL~\cite{Kwon:1981ua},
G\"osgen~\cite{Zacek:1986cu}, SRP~\cite{Greenwood:1996pb}, and
ROVNO88~\cite{Afonin:1988gx}  showed a deficit. In the framework
of three flavour oscillations, these reactor short-baseline experiments
(RSBL) were not sensitive to oscillations, but at the time played an
important role in constraining the unoscillated reactor neutrino flux.
So, they could be used as an alternative to  theoretically
calculated reactor fluxes.

The dependence of these early determinations of $\theta_{13}$ on the
reactor flux modeling is illustrated in Figure~\ref{fig:react-t13}.
The upper panels contain the $\Delta\chi^2$ from Chooz, Palo Verde,
Double Chooz, Daya Bay, and RENO as a function of $\theta_{13}$ for
different choices for the reactor fluxes.  The upper-left panel shows that, when the fluxes from~\cite{Huber:2011wv} had been
employed and   RSBL reactor experiments had not been included in the fit,
all experiments, including  Chooz and Palo Verde,
preferred $\theta_{13} > 0$. However, when the RSBL reactor
experiments had been added to the fit, such preference
vanished~\cite{Schwetz:2011qt}, and that happened independently of
whether  flux normalisation $f_\text{flux}$ was left as a free parameter
or not. This can also be inferred from the lower-left panel,
which shows  contours in the ($\theta_{13}$,
$f_\text{flux}$) plane for  analysis of Chooz and Palo Verde
with and without the inclusion of RSBL data.
The central panels show the dependence of the determination of $\theta_{13}$
from  analysis of Double Chooz on the choice of reactor fluxes:
the best-fit value and statistical significance
of the nonzero $\theta_{13}$ signal in this experiment  significantly depended
on the reactor flux assumption. This was due to the lack of the near detector
in Double Chooz at the time.

In view of this, and in order to properly assess the impact of the reactor
anomaly on the allowed range of neutrino parameters in NuFIT 1.0,
the global analysis was performed under two extreme choices.
In the first choice (``Free fluxes + RSBL'' in
Figure~\ref{fig:compachis}) we left the normalisation of reactor
fluxes free, and included  data from RSBL experiments.  In the second
option (``Huber''), we did not include the RSBL data,
and assume reactor fluxes and uncertainties as predicted in~\cite{Huber:2011wv}. The left panels of
Figure~\ref{fig:compachis} show that this choice resulted in an additional
uncertainty of about $1\sigma$ on various observables.

Being equipped with a near detector, the determination of
$\theta_{13}$ from Daya Bay and RENO was unaffected by the reactor
anomaly. As their statistics increased, and with the entrance in
operation of the Double Chooz near detector, the impact of the reactor
flux normalisation uncertainty steadily decreased,  reduced to
$\sim 0.5\sigma$ in  NuFIT~2.0 analysis, and becoming essentially
irrelevant in NuFIT~3.0.

In what respects the analysis of KamLAND long baseline reactor data,
since NuFIT~4.0 we have been relying on the precise reconstruction of
the reactor neutrino fluxes (both overall normalisation and energy
spectrum) provided by the Daya Bay near detectors~\cite{An:2016srz},
which renders also the KamLAND analysis largely independent of the
reactor anomaly. As a side effect, this change in the KamLAND reactor
flux model is responsible for the slight increase (from 14\% to 16\%)
of the $\Dmq_{21}$ uncertainty which can be observed in
Equation~\eqref{eq:compaprec}.

\begin{figure}[H]
  \includegraphics[width=\linewidth]{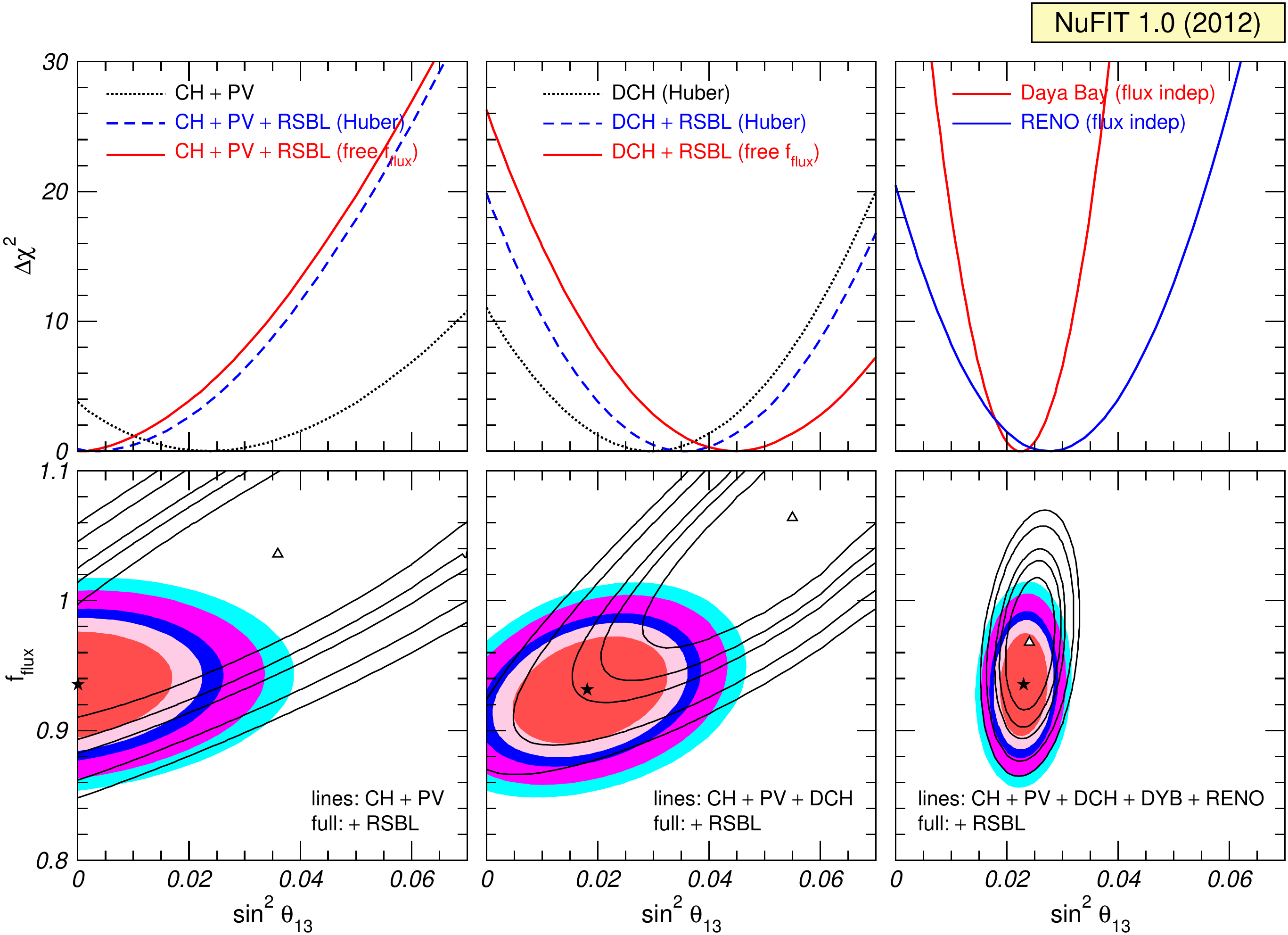}
  \caption{Dependence on  reactor flux normalisation choice in
    NuFIT~1.0. {\bf (upper)} Dependence of $\Delta\chi^2$ on
    $\sin^2\theta_{13}$ (for fix value $\Dmq_{31}=2.47\times
    10^{-3}~\eVq$) for the set of reactor experiments included in the
    analysis and three different assumptions on the fluxes as labeled
    in the figure.  {\bf (lower)} Confidence level contours in the
    plane of $\sin^2\theta_{13}$ and  flux normalisation
    $f_\text{flux}$. Full regions and lines correspond to analysis
    with and without including the RSBL experiments, respectively.
    Figure adapted from~\recite{nufit}.}
  \label{fig:react-t13}
\end{figure}

\subsection{Status of $\Dmq_{21}$ in Solar Experiments versus KamLAND}
\label{sec:solkam-dm12}

Analyses of the solar experiments and of KamLAND give the dominant
contribution to the determination of $\Dmq_{21}$ and $\theta_{12}$.
Starting with NuFIT~2.0, and as illustrated in the upper panels in
Figure~\ref{fig:sun-tension},  results of  global analyses showed
a value of $\Dmq_{21}$ preferred by KamLAND, which was somewhat higher
than the value favoured by solar neutrino experiments.  This tension
arose from a combination of two effects that did not 
significantly change till 2020:
\begin{itemize}
\item the observed $^8\text{B}$ spectrum at SNO, SK, and
  Borexino showed no clear evidence of the low-energy  turn-up,
   which is predicted to occur in the standard
  LMA-MSW~\cite{Wolfenstein:1977ue, Mikheev:1986gs} solution for the
  value of $\Dmq_{21}$ that fits KamLAND best.

\item Super-Kamiokande observed a day--night asymmetry
  that was larger
  than expected for the $\Dmq_{21}$ value preferred by KamLAND for
  which Earth matter effects are very small.

\end{itemize}

These effects resulted in the best-fit value of $\Dmq_{21}$ of KamLAND
in the NuFIT~2.0 fit lying at the boundary of the allowed $2\sigma$ 
range of the solar neutrino analysis, as seen in the upper panels in
Figure~\ref{fig:sun-tension}.  The tension was maintained with the
increased statistics from SK-IV included in NuFIT~3.0 and the change
in the reactor flux normalisation used in the KamLAND analysis since
NuFIT~4.0.

\begin{figure}[H]
  \includegraphics[width=0.8\linewidth]{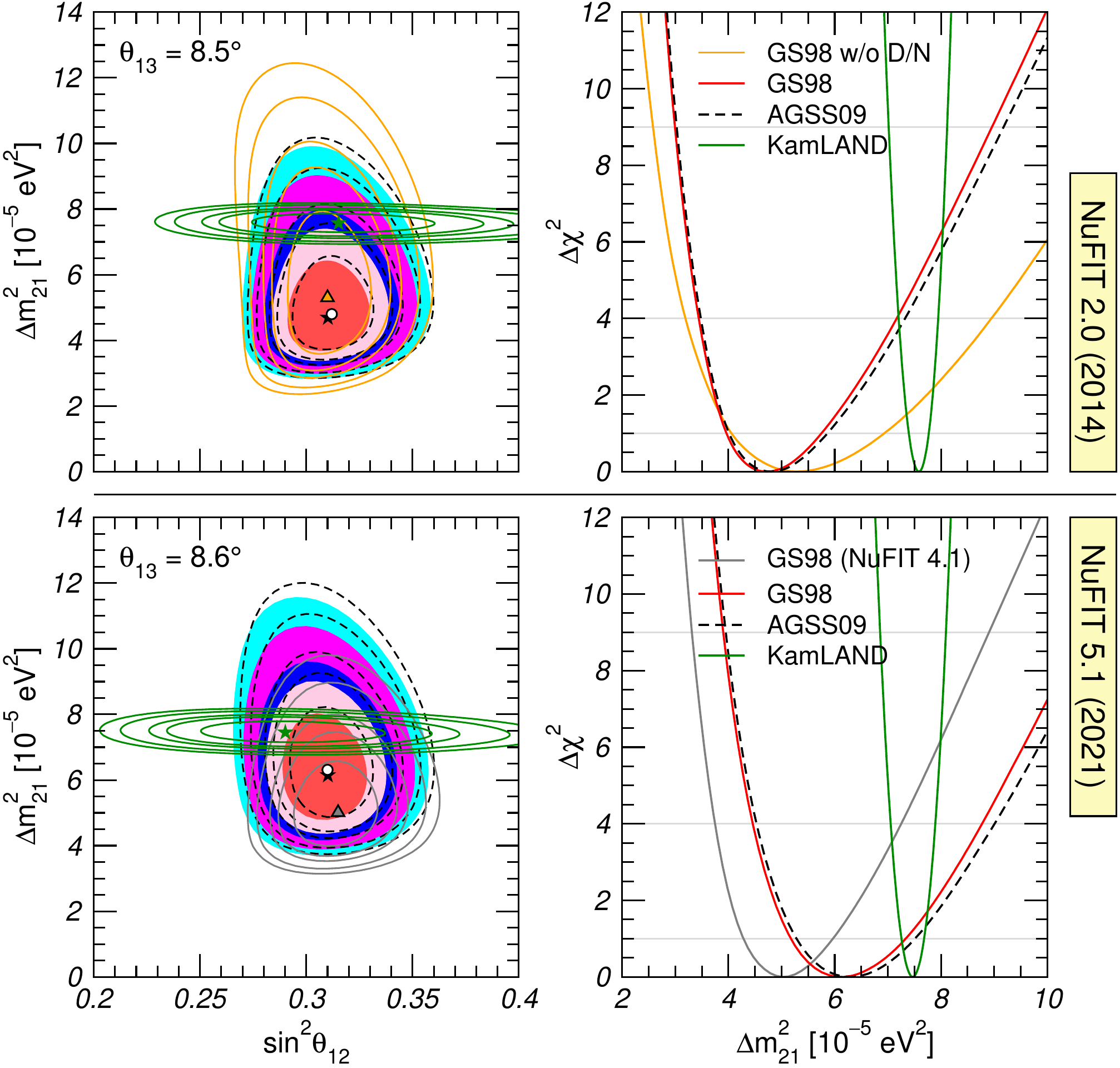}
  \caption{\textbf{(upper-left)} NuFIT~2.0 allowed parameter regions
    (at $1\sigma$, 90\%, $2\sigma$, 99\% and $3\sigma$ CL for
    2~d.o.f.) from  combined analysis of solar data for  GS98
    model (full regions with best fit marked by black star) and 
    AGSS09 model (dashed void contours with best fit marked by a white
    dot), and analysis of KamLAND data (solid green
    contours with best fit marked by a green star) for fixed
    $\theta_{13}=8.5^\circ$. Orange contours are the
    results of a global analysis for the GS98 model but without
    including  day--night information from SK. \textbf{(upper-right)}
    NuFIT~2.0 $\Delta\chi^2$ dependence on $\Dmq_{21}$ for the same
    four analyses after marginalising over $\theta_{12}$.
    \textbf{(lower-left)} Same as upper-left but for NuFIT~5.1 and
    fixed $\theta_{13}=8.6$.  Previous results of  global analysis for  GS98 model in
    NuFIT~4.0  shown as orange contours~\recite{Esteban:2018azc}.  \textbf{(lower-right)}
    $\Delta\chi^2$ dependence on $\Dmq_{21}$ for  same four
    analyses as the lower-left panel after marginalising over
    $\theta_{12}$.  Figure adapted from~\recite{nufit}.}
  \label{fig:sun-tension}
\end{figure}

The tension was resolved with the latest SK4 2970-day results
included in NuFIT~5.0, which were presented at the Neutrino2020
conference~\cite{SK:nu2020} in the form of their total energy spectrum,
which showed a slightly more pronounced turn-up in the low-energy part, and the updated day--night asymmetry
\begin{equation}
  \label{eq:newadn}
  A_\text{D/N}^\text{SK4-2970} = (-2.1\pm 1.1)\% \,.
\end{equation}
which was lower than the previously reported value
$A_\text{D/N,SK4-2055} =
  [-3.1\pm 1.6(\text{stat.}) \pm 1.4(\text{syst.})]\%$.

The impact of these new data is displayed in the lower panels of
Figure~\ref{fig:sun-tension}.  The tension between 
best fit $\Dmq_{21}$ of KamLAND and that of the solar results
decreased, and the preferred $\Dmq_{21}$ value from KamLAND lay at
$\Delta\chi^2_\text{solar} = 1.3$ (corresponding to $1.1\sigma$).
This decrease in  tension was due to both the smaller day--night
asymmetry (which lowered $\Delta\chi^2_\text{solar}$ of the KamLAND
best fit $\Dmq_{21}$ by $2.4$ units) and the slightly more pronounced
turn-up in the low-energy part of the spectrum (which lowered it by
one extra unit).

Lastly, in order to quantify the independence of these results on
the details of the solar modelling,  solar neutrino analysis in NuFIT
was performed  for the two versions of the Standard Solar Model,
namely, the GS98 and the AGSS09 models, which 
emerged as a consequence of the new determination of the abundances of
heavy elements because no viable Solar Standard Model could be constructed
that could accommodate these new abundances   with
 observed helioseismological data.
Consequently, two different sets of models were constructed that are in
better accordance with one or the other~\cite{Vinyoles:2016djt,Bergstrom:2016cbh}. From the point of view of  solar neutrino analysis,
the existence of these two SSM variants  is relevant because
they differ in the predicted neutrino fluxes, in particular those generated
in the CNO cycle. This introduces a possible source of theoretical uncertainty
in the determination of  relevant oscillation parameters.
In NuFIT, we quantify this possible uncertainty by performing  analysis
with both models.  
Figure~\ref{fig:sun-tension} shows that the determination of $\Dmq_{21}$ and $\theta_{12}$ is
extremely robust over these variations on the modelling of the Sun.

\subsection{Inclusion of Super-Kamiokande Atmospheric Neutrino Data}
\label{sec:skatm}

Atmospheric neutrinos are produced by the interaction of cosmic rays
on the top of  Earth's atmosphere. In the subsequent hadronic
cascades, both $\nu_e$ and $\nu_\mu$, and $\bar\nu_e$ and
$\bar\nu_\mu$ are produced with a broad range of energies. Furthermore,
atmospheric neutrinos are produced in all possible
directions. Therefore, at any detector positioned on  Earth, a good fraction
of  events generated by the interaction of these neutrinos come
from neutrinos that have traveled through  Earth.
For all these reasons, atmospheric neutrinos
constitute a powerful tool to study the evolution of neutrino flavour
in their propagation.

In the context of three flavour oscillations,  atmospheric neutrino
data show that the dominant oscillation channel of atmospheric
neutrinos is $\nu_\mu\to \nu_\tau$, which in the standard convention
described in Section~\ref{sec:frame} is driven by $|\Dmq_{31}|$ and with
the amplitude controlled by $\theta_{23}$.  In principle, the flavour,
and neutrino and antineutrino composition of  atmospheric neutrino
fluxes, together with a wide range of covered baselines, open up the
possibility of sensitivity to subleading oscillation modes, driven by
$\Dmq_{21}$ and/or $\theta_{13}$, especially in  light of the
not-too-small value of $\theta_{13}$. In particular, they 
could provide relevant information on the octant of
$\theta_{23}$,  the value of $\dCP$, and the ordering of the neutrino
mass spectrum.

In NuFIT~1.0 and NuFIT~2.0, we performed our own analysis of 
Super-Kamiokande atmospheric neutrino data for phases SK1--4.  The
analysis was based on classical data samples---sub-GeV and
multi-GeV $e$-like and $\mu$-like events, and
partially contained, stopping, and through-going muons---which
accounted for a total of 70 data points, and for which one could
perform a reasonably accurate simulation using the information
provided by the collaboration. The implications of our last SK
analysis of such kind in the global picture is shown in
Figure~\ref{fig:v20.chisq-hier}:  the impact on both the
ordering and the determination of $\Delta\chi^2$ was modest.

Around that time, Super-Kamiokande started developing a dedicated
analytical methodology  for constructing $\nu_e + \bar\nu_e$
enriched atmospheric neutrino samples and further classifying them
into $\nu_e$-like and $\bar\nu_e$-like subsamples. With those, they
seemed to have succeeded at increasing their sensitivity to the subleading
effects.  With the limited information available
outside of the collaboration, it was not possible to reproduce  key
elements driving the main dependence on these subdominant oscillation
effects.  Consequently, our own simulation of  SK atmospheric data
fell short at this task, and since NuFIT-3.0 they have been removed from our
global analysis.

\begin{figure}[H]
  \includegraphics[width=\linewidth]{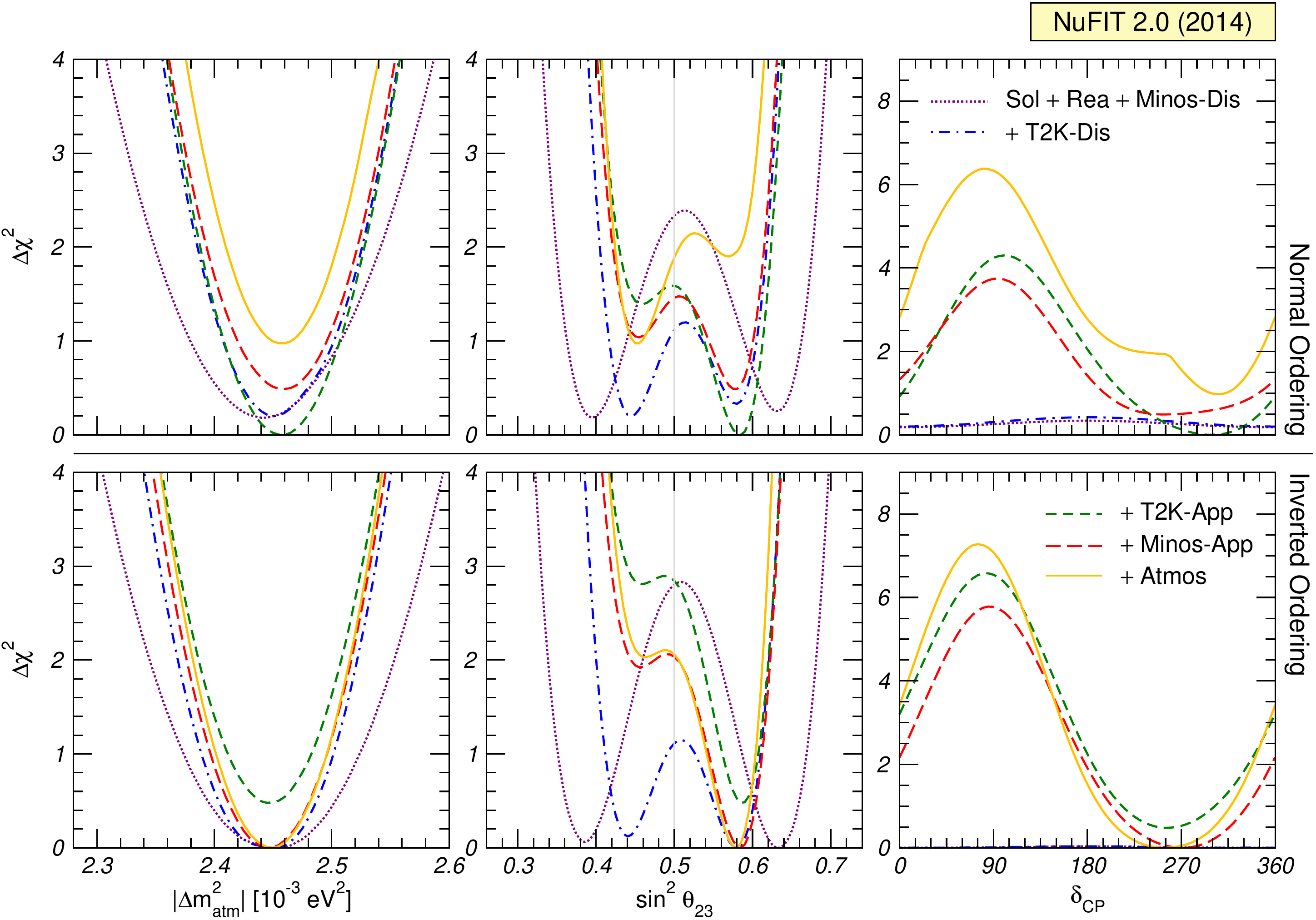}
  \caption{Contribution of different sets of experimental data
    included in NuFIT~2.0 to the determination of  mass ordering,
     octant of $\theta_{23}$, and  CP violating phase. Left
    (right) panels are for IO (NO). Atmos is
    our analysis of SK1--4 (including SK4 1775-day) combined data.
    Figure adapted from~\recite{nufit}.}
  \label{fig:v20.chisq-hier}
\end{figure}

In 2017, the Super-Kamiokande collaboration started to publish results
obtained with this method~\cite{Abe:2017aap}  providing the
corresponding tabulated $\chi^2$ map~\cite{SKatm:data2018} as a
function of the four relevant parameters $\Dmq_{3\ell}$,
$\theta_{23}$, $\theta_{13}$, and $\dCP$. Such a table could be added to
the $\chi^2$ of our global analysis to address the impact of their
data in the global picture. This was performed in NuFIT~4.X and NuFIT~5.0
versions.  Super-Kamiokande   publicised an updated table
with a slight increase in  exposure~\cite{SKatm:data2020}.  The
effect of adding that information was  included in our last
analysis, NuFIT~5.1, and  is shown as dashed curves in the right
panels of Figure~\ref{fig:compachis}; see also Table~\ref{tab:bfranges}). The addition
of the SK-atm table to the latest analysis resulted in an increase of the favouring
of NO and in the significance of CP violation, and a change in the
favoured octant of $\theta_{23}$.

However,  this procedure of ``blindly adding''
the $\chi^2$ table, as provided by the experimental collaboration, is
not optimal as it defeats the purpose of the global phenomenological
analysis, whose aim is both reproducing and combining different data
samples under a consistent set of assumptions on the theoretical
uncertainties, as well as exploring the implication of the results in
extended scenarios.

\subsection{$\theta_{23}$, $\dCP$ and Mass Ordering from  LBL Accelerator and MBL Reactor Experiments}
\label{sec:lblreac}

From the point of view of  data included in  analysis, the most
important variation over the last decade was in LBL
accelerator and MBL reactor experiments.

The data included in NuFIT~1.0 for LBL experiments comprised the
spectrum of $\nu_\mu$ disappearance events of K2K~\cite{Ahn:2006zza},
both $\nu_\mu$ ($\bar\nu_\nu$) disappearance and $\nu_e$ ($\bar\nu_e$)
appearance spectra in MINOS with $10.8~(3.36) \times 10^{20}$ protons
on target (pot)~\cite{minos:nu2012}, and the results from T2K $\nu_e$
appearance and $\nu_\mu$ disappearance data of phases 1--3 ($3.01\times
10^{20}$ pot~\cite{t2k:ichep2012}) and phases 1--2 ($1.43\times
10^{20}$ pot~\cite{Abe:2012gx, t2k:nu2012}), respectively.  NOvA data
were first available in NuFIT~3.0.  NuFIT~5.X includes the latest
results from T2K corresponding to $19.7\times 10^{20}$ pot
($16.3\times 10^{20}$ pot) $\nu$ ($\bar\nu$) spectra in both $\nu_\mu$
($\bar\nu_\mu$) disappearance and $\nu_e$ ($\bar\nu_e$) appearance
data~\cite{T2K:nu2020}, as well as NOvA data corresponding to
$13.6\times 10^{20}$ pot ($12.5\times 10^{20}$ pot) $\nu$ ($\bar\nu$)
spectra in both $\nu_\mu$ ($\bar\nu_\mu$) disappearance and $\nu_e$
($\bar\nu_e$) appearance~\cite{NOvA:nu2020}.     Regarding MBL
reactor data, NuFIT~1.0 included  results of 126 live days of Daya
Bay~\cite{dayabay:nu2012} and 229 days of data taking of
RENO~\cite{Ahn:2012nd} in the form of total event rates in the near
and far detectors, together with the initial spectrum from Double
Chooz far detector with 227.9 days live time~\cite{Abe:2012uy,
  dc:nu2012}.  In NuFIT~5.X we account for the results of the 1958-day
EH2/EH1 and EH3/EH1 spectral ratios from Daya Bay~\cite{Adey:2018zwh},
the 2908-day FD/ND spectral ratio from RENO~\cite{RENO:nu2020}, and
the Double Chooz FD/ND spectral ratio with 1276-day (FD) and 587-day
(ND) exposures~\cite{DoubleC:nu2020}.

The increase in available data  in both types of experiments and  their complementarity   played the leading role
in the observation of subdominant effects associated to $\dCP$,  
neutrino mass ordering, and  the octant of $\theta_{23}$, with
hints of favoured values and their statistical significance
changing in time.  We illustrate this in Figure~\ref{fig:lblreac} which shows
 $\Delta\chi^2$ profiles as a function of these three
parameters in the last two NuFIT analyses.

\end{paracol}
\nointerlineskip
\begin{figure}[H]
\widefigure
  \includegraphics[width=\linewidth]{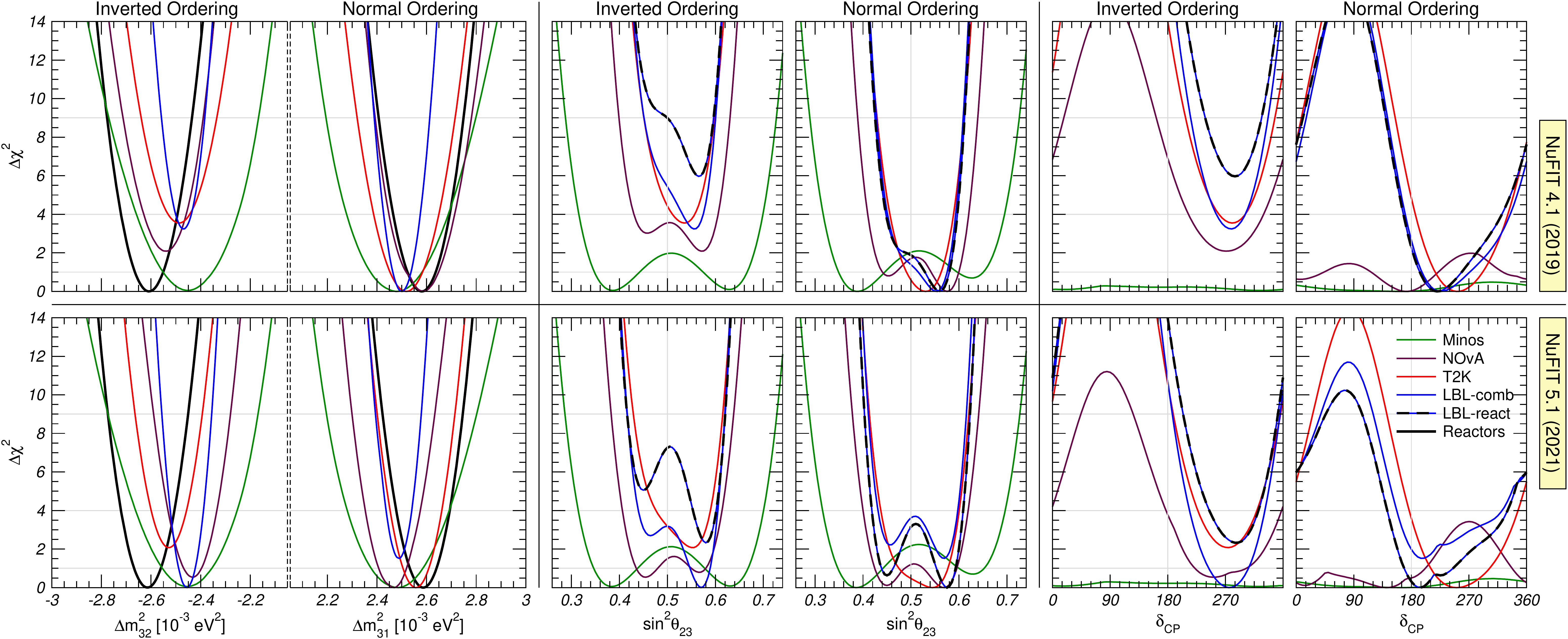}
  \caption{{$\Delta\chi^2$-profiles} as a function of $\Dmq_{3\ell}$
    (\textbf{left}), $\sin^2\theta_{23}$ (\textbf{center}) and $\dCP$ (\textbf{right}) for
    different LBL and reactor data sets and their combination obtained
    in NuFIT~4.1 and NuFIT~5.1.  For all curves we have fixed
    $\sin^2\theta_{13}=0.0224$ as well as the solar parameters and
    minimised with respect to the other undisplayed parameters.
    $\Delta\chi^2$ is shown with respect to the best-fit mass ordering
    for each curve. Figure adapted from~\recite{nufit}.}
  \label{fig:lblreac}
\end{figure}
\begin{paracol}{2}
\switchcolumn

Qualitatively, the most relevant effects can be understood in terms of
approximate expressions for the relevant oscillation probabilities. In
particular, the $\nu_\mu$ survival probability is given to good
accuracy by~\cite{Okamura:2004if, Nunokawa:2005nx}
\begin{equation}
  P_{\mu\mu} \approx 1 - \sin^22\theta_{\mu\mu} \sin^2\frac{\Dmq_{\mu\mu} L}{4E_\nu} \,,
\end{equation}
where $L$ is the baseline, $E_\nu$ is the neutrino energy, and
\begin{align}
  \sin^2\theta_{\mu\mu}
  &= \cos^2\theta_{13} \sin^2\theta_{23} \,,
  \\
  \label{eq:Dmqmm}
  \Dmq_{\mu\mu}
  &= \sin^2\theta_{12}\, \Dmq_{31} + \cos^2\theta_{12}\, \Dmq_{32}
  + \cos\dCP \sin\theta_{13}\sin 2\theta_{12} \tan\theta_{23}\, \Dmq_{21} \,.
\end{align}
The $\nu_e$ survival probability relevant for reactor
experiments with MBL can be approximated as~\cite{Nunokawa:2005nx,
  Minakata:2006gq}:
\begin{equation}
  P_{ee} \approx 1 - \sin^22\theta_{13} \sin^2\frac{\Dmq_{ee} L}{4E_\nu} \,,
\end{equation}
where
\begin{equation}
  \label{eq:Dmqee}
  \Dmq_{ee} = \cos^2\theta_{12}\, \Dmq_{31} + \sin^2\theta_{12}\, \Dmq_{32} \,.
\end{equation}
{Hence,} the determination of the oscillation frequencies in $\nu_\mu$
and $\nu_e$ disappearance experiments provides two independent
measurements of the parameter $|\Dmq_{3\ell}|$, which already in
NuFIT~2.0 were of similar accuracy and therefore allowed for a
consistency test of the $3\nu$ scenario. Furthermore, as precision
increased the comparison of both oscillation frequencies started
offering relevant information on the sign of $\Dmq_{3\ell}$,
{i.e.}, contributing to the present sensitivity to the mass
ordering.

For the $\nu_e$ appearance results in T2K and NOvA, following~\cite{Elevant:2015ska, Esteban:2018azc}, qualitative
understanding can be obtained by expanding the appearance oscillation
probability in the small parameters $\sin\theta_{13}$, $\Dmq_{21}
L/E_\nu$, and the matter potential term $A \equiv | 2E_\nu V /
\Dmq_{3\ell}|$ ($L$ is the baseline, $E_\nu$ the neutrino energy and
$V$ the effective matter potential):
\begin{align}
  P_{\nu_\mu\to\nu_e}
  &\approx 4 s_{13}^2s_{23}^2(1+2\,s\,A) - C \sin\dCP(1+s\,A) \,,
  \\
  P_{\bar\nu_\mu\to\bar\nu_e}
  &\approx 4 s_{13}^2s_{23}^2(1-2\,s\,A) + C \sin\dCP(1-s\,A) \,.
\end{align}
with $s_{ij} \equiv \sin\theta_{ij}$ and
\begin{equation}
  C \equiv \frac{\Dmq_{21}L}{4E_\nu}
  \sin2\theta_{12}\sin2\theta_{13}\sin2\theta_{23}\,,
  \qquad
  s \equiv \mathop{\mathrm{sign}}(\Dmq_{3\ell}) \,,
\end{equation}
and we  used $|\Dmq_{3\ell}|\, L/4E_\nu \approx \pi/2$ for both
T2K and NOvA.  Using the average value of  Earth's crust matter
density,  neutrinos are found with mean energy at T2K $A
\approx 0.05$, whereas for NOvA, the approximation works best with 
an empirical value of $A=0.1$.  Under the approximation  that
the total number of appearance events observed in T2K and NOvA is
proportional to the oscillation probability one can write
\begin{align}
  \label{eq:Nnu}
  N_{\nu_e}
  &\approx \mathcal{N}_\nu
  \left[ 2 s_{23}^2(1+2\,s\,A) - C' \sin\dCP(1+s\,A) \right] \,,
  \\
  \label{eq:Nan}
  N_{\bar\nu_e} &
  \approx \mathcal{N}_{\bar\nu}
  \left[ 2 s_{23}^2(1-2\,s\,A) + C' \sin\dCP(1-s\,A) \right] \,.
\end{align}
{When} all the well-determined parameters $\theta_{13}$, $\theta_{12}$,
$\Dmq_{21}$, $|\Dmq_{3\ell}|$ are set to their global best fit values,
one gets $C' \approx 0.28$ almost independently of the value of
$\theta_{23}$.  The normalisation constants
$\mathcal{N}_{\nu,\bar\nu}$ can be calculated from the total number of
events in the different appearance samples.

For the last few years, T2K data have been favouring a ratio of
observed/expected events larger than 1 for neutrinos and smaller than
1 for anti-neutrinos. The expressions in Equations~\eqref{eq:Nnu}
and~\eqref{eq:Nan}  imply that the square-bracket term in
Equation~\eqref{eq:Nnu} had to be enhanced, and the one in
Equation~\eqref{eq:Nan} had to be suppressed. With $\theta_{13}$ fixed by
reactor experiments, this could be achieved by choosing NO and $\dCP
\simeq 3\pi/2$. This  is the driving factor for the hints in
favour of NO and maximal CP violation since NuFIT 3.0.  NOvA neutrino data    indicate towards ratios
closer to 1, which can be accommodated by either (NO, $\dCP \simeq
\pi/2$) or (IO, $\dCP \simeq 3\pi/2$).  This behaviour is consistent
with NOvA antineutrinos, but the NO option is somewhat in tension
with T2K. This small tension between T2K and NOVA  resulted in
variations in the favoured ordering in  combined LBL analysis and the favoured octant of $\theta_{23}$ and value of $\dCP$ in
NuFIT 4.X and NuFIT 5.X.

On the other hand, with  respect to the complementary accelerator/reactor 
determination of the oscillation frequencies in $\nu_\mu$ and $\nu_e$
disappearance experiments,  they have been consistently indicating
towards a better agreement for NO than that for IO, albeit within the
limited statistical significance of the effect.

\section{Conclusions}

Over the last two decades, neutrino oscillation experiments have
provided us with undoubted evidence that neutrinos have mass, and that
the lepton flavours mix in the charge current weak interaction of those
massive states.  Those observations, which cannot be explained within
the Standard Model, represent our only laboratory evidence of physics
beyond the Standard Model.

The determination of the flavour structure of the lepton sector at low
energies is, at this point, our only source of information to
understand the underlying BSM dynamics responsible for these
observations, and it is therefore fundamental to ultimately establish
the \emph{{New} Standard Model}.

The task is at the hands of phenomenological groups.  NuFIT was
formed in this context about 10 years ago as a fluid collaboration. 
Since then, it has provided updated results from the global analysis of
neutrino oscillation measurements.  The NuFIT analysis is performed
in the framework of the Standard Model extended with three massive
neutrinos, which is currently the minimal scenario capable
of accommodating all oscillation results that were  robustly established.  In this
contribution, we summarised some results obtained by NuFIT over
these decade, in particular describing those issues which were solved
by new data and those which are still pending.

\vspace{6pt}
\authorcontributions{{All authors have contributed equally to this article}}

\funding{This work is supported by Spanish grants PID2019-105614GB-C21
  and PID2019-110058GB-C21 financed by MCIN/AEI/10.13039/501100011033,
  by USA-NSF grant PHY-1915093, and by AGAUR (Generalitat de
  Catalunya) grant 2017-SGR-929. The authors acknowledge the support
  of European ITN grant H2020-MSCA-ITN-2019//860881-HIDDeN and of the
  Spanish Agencia Estatal de Investigación through the grant ``IFT
  Centro de Excelencia Severo Ochoa SEV-2016-0597''.}
  
  \institutionalreview{{Not applicable}}

\informedconsent{{Not applicable}}



\conflictsofinterest{{The authors declare no conflict of interest}} 

\end{paracol}

\reftitle{References}


\end{document}